\documentclass[letterpaper, 10 pt, conference]{ieeeconf}  

\usepackage{microtype}
\usepackage{graphicx}
\usepackage{subfigure}
\usepackage{booktabs}
\usepackage{epsfig}
\usepackage[labelfont=bf]{caption}
\usepackage{indentfirst}
\usepackage{amsmath}
\usepackage{amssymb}
\usepackage{mathtools}
\usepackage{footmisc}
\usepackage{xcolor}
\usepackage[pagebackref,breaklinks,colorlinks]{hyperref}
\usepackage[capitalize,noabbrev]{cleveref}
\usepackage{float}
\usepackage[normalem]{ulem}

\IEEEoverridecommandlockouts                
\title{\LARGE \bf
Enhancing the Performance of Multi-Vehicle Navigation in Unstructured Environments using Hard Sample Mining}

\author{Yining Ma$^{1,2,3}$, Ang Li$^{1}$, Qadeer Khan$^{1,2}$ and Daniel Cremers$^{1,2}$
\thanks{$^{1}$ Computer Vision Group, TU Munich.  }%
\thanks{$^{2}$ Munich Center for Machine Learning (MCML).}
\thanks{$^{3}$ The Konrad Zuse School of Excellence in Learning and Intelligent Systems (ELIZA).}
}

\begin{document}

\maketitle
\thispagestyle{empty}
\pagestyle{empty}

\begin{abstract}

Contemporary research in autonomous driving has demonstrated tremendous potential in emulating the traits of human driving. However, they primarily cater to areas with well-built road infrastructure and appropriate traffic management systems. Therefore, in the absence of traffic signals or in unstructured environments, these self-driving algorithms are expected to fail. This paper proposes a strategy for autonomously navigating multiple vehicles in close proximity to their desired destinations without traffic rules in unstructured environments.

Graphical Neural Networks (GNNs) have demonstrated good utility for this task of multi-vehicle control. Among the different alternatives for training GNNs, supervised methods have proven to be the most data-efficient, albeit requiring ground truth labels. However, these labels may not always be available, particularly in unstructured environments without traffic regulations. Therefore, a tedious optimization process may be required to determine them while ensuring that the vehicles reach their desired destination and do not collide with each other or any obstacles. Therefore, in order to expedite the training process, it is essential to reduce the optimization time and select only those samples for labeling that add the most value to the training. 

In this paper, we propose a warm start method that first uses a pre-trained model trained on a simpler subset of data. Inference is then done on more complicated scenarios to determine the hard samples wherein the model faces the greatest predicament. This is measured by the difficulty vehicles encounter in reaching their desired destination without collision. Experimental results demonstrate that mining for hard samples in this manner reduces the requirement for supervised training data by 10 fold. Moreover, we also use the predictions of this simpler pre-trained model to initialize the optimization process, resulting in a further speedup of up to 1.8 times. Videos and code can be found on the project page: \url{https://yininghase.github.io/multiagent-collision-mining/}. 

\end{abstract}

\section{INTRODUCTION}\label{sec:introduction}

We have already seen the evolution of driver-less taxis being used to conveniently transport people between places \cite{kim2017autonomous}. However, the deployment of such algorithms is mostly constrained to areas with well-built road infrastructure and explicit traffic regulations. How are such self-driving algorithms supposed to react where lane markings or even paved roads are absent from the onset? Such learning-based self-driving algorithms are expected to fail due to domain shift \cite{ross2010efficient,pmlr-v87-wenzel18a,pomerleau1988alvinn}. This is particularly true of many places in the developing world where road infrastructure is sparse \cite{mejia2020delay} and appropriate traffic management systems may not be ubiquitous \cite{saha2020automated}. This leads to drivers having to negotiate right-of-way with other road participants in real time to navigate through to their desired destination while avoiding collisions. The absence of traffic management in unstructured environments necessitates extra attention from the driver, thereby exacerbating the risk of accidents, particularly caused by human factors \cite{bucsuhazy2020human}.  Therefore, allowing the vehicle to autonomously navigate in such taxing scenarios relieves the burden from the human, thereby potentially reducing the risk of crashes \cite{devcountries} caused by lack of attention.  However, most contemporary research in autonomous driving has not been centered towards such scenarios but rather on structured road environments \cite{zhang2022learning,chitta2022transfuser,chen2022learning,khan2023learning}. On the other hand, in this paper, we propose a framework for autonomously navigating multiple vehicles in the absence of traffic signals in unstructured environments. Such environments can arise not only in areas with limited road infrastructure but also for e.g., in large warehouses where multiple robot agents must carry items from one place to another.

Deep learning methods employing graphical neural network architectures have emerged as a viable option for multi agent trajectory prediction and control \cite{10186793,10018862,9564929,ZHU2024127220,9827155}. However, such methods are mainly confined to environments with well-defined road markings \& infrastructure. \cite{9551450,multiagent2023} do cater to unstructured scenes. However, the former is limited to parking garages, where the paths are still evident, while the latter can only handle scenarios with limited vehicles. \cite{li2020graph} conducts path planning of multiple agents. However, the control space is discrete, thereby limiting the number of actions that can be executed. In contrast, the action space of our model is continuous, thereby giving a wide range of steering maneuvers.  \

\begin{figure*}[!h]
\centering
\includegraphics[width=1.6\columnwidth]{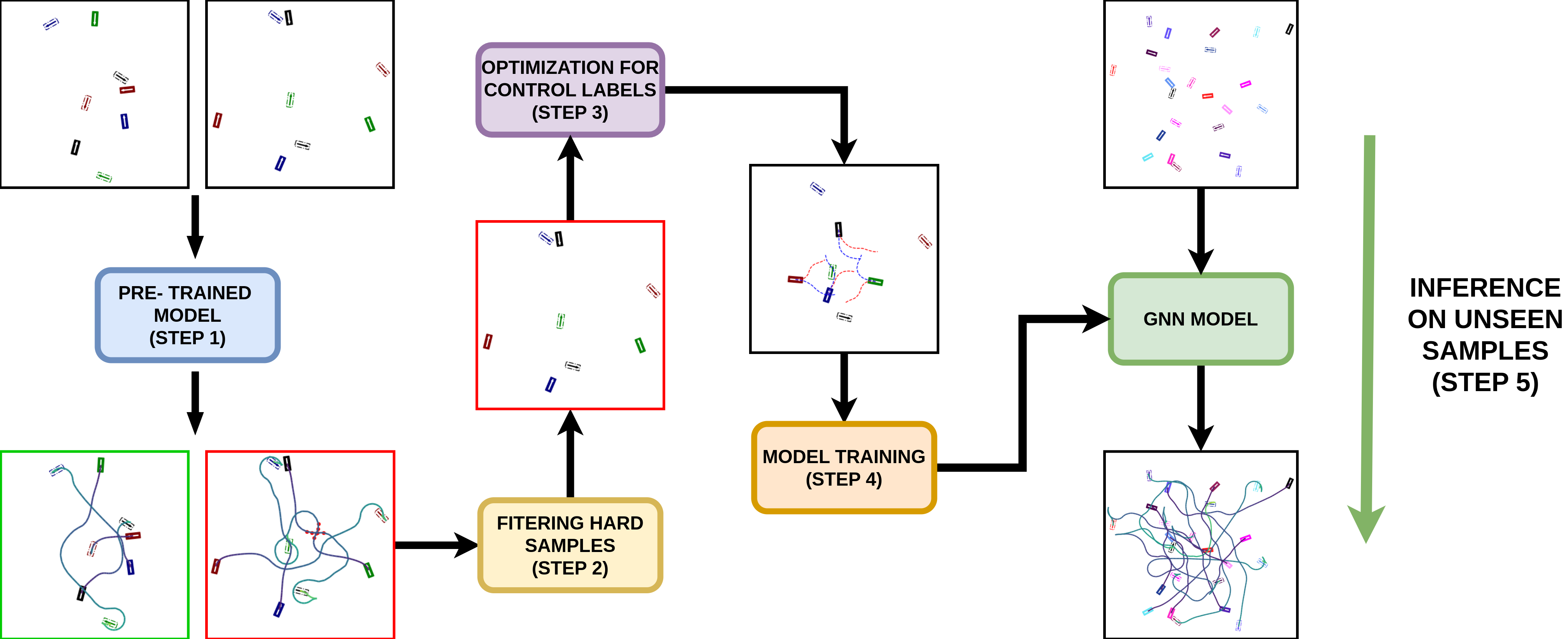}
\caption{Shows the overview of our framework. \textbf{Step 1:} A pre-trained model trained on a simpler subset of data is first executed on multiple complicated scenarios to infer the corresponding trajectories. Scenarios producing no collision in their trajectories are depicted by a green border, while others are shown in red. The points of collision are depicted by red dots. For simplicity, we have only shown two out of the many scenarios. \textbf{Step 2}: Samples with no or minimal collisions are filtered out, leaving only the examples that were difficult for the pre-trained model to predict the correct control labels. \textbf{Step 3}: These remaining difficult/hard samples are passed to an optimization algorithm to produce the corresponding control labels (Shown in dotted red color). The dotted blue lines are the predictions from the pre-trained model that are used for warm-starting the iterative optimization process. \textbf{Step 4:} This newly labeled additional data is used to train a Graphical Neural Network (GNN) model, to improve performance. \textbf{Step 5:} Inference is done on unseen samples that can have even more vehicles than in the training set. Note that the black arrows depict the flow of the training process, while the green arrow shows the inference process. Out of the entire pipeline, only Step 5 is needed at test time.
 }
\label{fig:overview}
\vspace*{-5mm}
\end{figure*}

Supervised learning tends to be used for training the aforementioned models for multi-agent interactions. Other works have used reinforcement learning (RL) as an alternative for  multi-agent path planning and control \cite{lowe2017multi,sartoretti2019primal,haworth2020deep}.  However, \cite{lowe2017multi} models the vehicles as particles and does not consider their kinematics, \cite{sartoretti2019primal} only considers a discrete action space in a constrained grid-world, while \cite{haworth2020deep} caters only to humanoid robots.  Another point of concern in RL is safety during the learning phase. As opposed to supervised learning, RL agents are not explicitly informed on how to perform a specific action. Rather, they choose actions in order to explore the environment with some inherent randomness. The viability of the action is measured in terms of a reward function \cite{kiran2021deep,sutton2018reinforcement}. This exploratory nature of the agent in RL may lead to risky driving behaviour resulting in accidents when training. \cite{kendall2019learning} had a safety driver to override the controls of the RL agent whenever the car deviates from the driving path. Moreover, reinforcement learning tends to be more sample inefficient \cite{8686348,electronics9091363}, particularly when compared with supervised learning techniques. 

On the other hand, an issue with most supervised learning methods is that their success primarily hinges upon the availability of tremendous amounts of annotated training data for supervised training. Obtaining annotations is generally an expensive and time-consuming process. Moreover, labeling all samples may not even be humanly possible in various domains, such as autonomous driving.  This is because data generation from various sources and sensor modalities is outpacing the limits of human labeling capacity.

An alternate would be to diligently mine for hard examples \cite{10.1007/978-3-030-58568-6_8,diagnostics11030518,liu2023hard} rather than exhaustively. Various strategies have been proposed for circumventing the need for annotating the entire data and focusing only on those samples where a trained deep learning model faces the most difficulty in making predictions. One such strategy, for e.g., is active learning, wherein humans are asked to label a much smaller subset of samples deemed difficult by the model. This labeled subset of difficult samples is eventually included in the training set to train and expand the predictive power of the model.  Such a strategy has been shown to have reduced human labeling effort in diverse applications \cite{siddiqui2020viewal,8814236,8606452}. 

However, one issue with the active learning process described above is that it requires a human in the loop to  label the samples deemed to be difficult for the model. However, not all tasks can optimally be labeled by a human. This is the case with attempting to navigate multiple vehicles where road infrastructure does not exist and traffic rules are not clearly defined. It is not clear how human should control all the vehicles simultaneously so that they optimally reach their desired destination as quickly as possible. Hence, training a model with sub-optimally labeled samples by a human would deteriorate performance. In fact, human labeling in this scenario is rather subjective, and two people may come up with different solutions for the same task. This problem is further exacerbated when multiple vehicles need to be navigated in an unstructured environment. This is because, in constrained environments such as on roads, clear rules exist for navigation. For e.g., all vehicles must drive on one side of the road, wait at a red traffic light, etc. Therefore, there is a greater probability that two persons following the rules may come up with the same solution. However, in unstructured environments, people may actually come up with different but conflicting solutions.

As a substitute to the counter-productive subjective labeling criteria, it is imperative to have an objective measure to arrive at the optimal labels which can be used to train a reliable model.  A viable alternative to determine the control labels for all vehicles is to optimize against a cost function such that each vehicle reaches its destination without collision. While this may provide an optimal solution, the optimization becomes increasingly slower as more vehicles are added to the environment \cite{9595509,multiagent2023}. The optimization can be expedited if the initial values are close to the optimal solution. But how can this be achieved? For this, we first pre-train a model on a smaller and simpler subset of data. Predictions made by this model on a more complicated dataset are then used as an initialization to warm start the optimization.  Therefore, when compared with other tasks such as classification and segmentation, for which the methodology of applying active learning/hard negative mining is well established, the task of multi-vehicle navigation appears to be more complicated. Therefore, this paper proposes a framework to tackle this task.

Figure \ref{fig:overview} explains our framework for multi-vehicle control. The framework does not rely on a human for labeling. It rather considers an optimization-based procedure as an objective measure for optimal labels. Furthermore, the requirement for training data is reduced by mining for difficult samples.
In this regard, our contributions are enumerated as follows:

\begin{enumerate}
    \item We demonstrate that mining hard samples for training yields overall better performance than just choosing data points with random sampling
    \item We also show that our approach of hard sample mining requires up to 10 times less training data to achieve equivalent or better performance than random sampling.
    \item The time to determine the control labels by running an optimization-based procedure is further reduced by up to 1.8 times using predictions from the pre-trained model as the initial values for the optimization.
    \item Training code \& corresponding videos can be found on the project page here:  \url{https://yininghase.github.io/multiagent-collision-mining/}
\end{enumerate}

\section{Framework}\label{sec:method}

The task we are trying to solve is as follows: Given are the $N_{vehicle}$  dynamic vehicles and $N_{obstacles}$ static obstacles in an unstructured environment. In the absence of any traffic management system (traffic light, sign, etc), we would like to predict the control command (steering angle $\varphi$ and pedal acceleration $p$) for each vehicle such that it can reach its desired destination without colliding with each other or with the static obstacles.

\subsection{Graphical Neural Network}\label{sec:GNN}
We use a graphical neural network (GNN)  to represent the scene. Therein, the nodes embed information about the agents that relay information to all other neighboring vehicles using message passing. GNNs provide an appropriate representation for this task since they allow invariance to isomorphism \cite{murphy2018janossy} and also to the number of vehicles due to shared network weights \cite{kipf2016semi}. This facilitates GNNs to perform inference for even more vehicles than for which it was initially trained, irrespective of the vehicle order. 

The input feature vector for each vehicle/obstacle node is given by $z \in \mathbb{R}^{8}$. This represents: the current location ($x$ and $y$), current orientation ($\theta$), current velocity $v$, target position ($\hat{x}$ and $\hat{y}$), target orientation ($\hat{\theta}$) and if the entity is a vehicle (0) or a circular obstacle (with radius $r$). In summary, 
$z_{vehicle} = [x, y, \theta, v, \hat{x}, \hat{y}, \hat{\theta}, 0]^{T}$, while, $z_{obstacle} = [x, y, \theta, 0, x, y, \theta, r]^{T}$. The choice of circular representation for the obstacle is elaborated further in Subsection \ref{sec:new_model_training}.

The graph $G$ is completed by adding edges between the nodes. For each vehicle node, an incoming edge is connected to every other vehicle \& obstacle node. Hence, the neighbours $j$ of vehicle $i$ in graph $G$ are  $N_{i} = \{ j | j = 0,1,2...,N_{vehicles}+N_{obstacle} \cap j \neq i \}$. These edges allow each vehicle to acquire information about the other entities using message passing. This information is then processed using a neural network with shared weights between the nodes. The output is then passed to successive GNN layers until we reach the final layer, which ultimately predicts an output $\in \mathbb{R}^{2}$ for the steering angle ($\varphi$) \& pedal acceleration ($p$).

\subsection{Mining for Hard Samples}\label{sec:hard_data_mining}

The purpose of mining for hard samples is two-fold:
\begin{enumerate}
    \item To improve the performance of the model wherein it is likely to face the most difficulty.
    \item To assess if the number of samples to be labeled can be reduced.
\end{enumerate}

In order to mine for the hard samples, we propose a warm start mechanism. First, we pre-train a model on a simpler training dataset with a maximum of 3 vehicles in the scene. We chose this number since optimizing for any larger number of vehicles considerably increases the optimization runtimes as depicted in Table \ref{table:statistic of optimization run time} ($3^{rd}$ column). The samples in this simple dataset are selected randomly. This pre-trained model is then used to run inference on more complicated scenarios \& with more vehicles in order to excavate out the hard/difficult samples. The difficulty of a sample is quantified by measuring the total number of potential collisions that are expected to occur divided by the total distance traveled by all vehicles in a particular scenario.  Scenarios with the highest values for this metric are chosen as the most useful candidates for which labels are to be determined and included in the training set. In the Experiments Section \ref{sec:experiments}, we show that mining for hard samples in this manner adds the most value to the training and reduces the labeling effort by 10-fold to achieve performance at par or even better than without hard sample mining. 
Once the hard samples for training have been identified, the next step is to determine their ground truth control labels. This is done using an optimization-based procedure described in Subsection \ref{sec:new_model_training}.

\subsection{Vehicle Kinematics}\label{sec:vehicle_kinematics}
The kinematics of the vehicles are simplified using the bicycle model \cite{wang2001trajectory}. It describes how
the equations of motion can be updated in time increments of $\Delta t$, assuming there is no slip condition:
\begin{equation}
\begin{split}
x_{t+1} & = x_{t} + v_{t} \cdot \cos(\theta_{t}) \cdot \Delta t \\ y_{t+1} & = y_{t} + v_{t} \cdot \sin(\theta_{t}) \cdot \Delta t \\ \theta_{t+1} & = \theta_{t} + v_{t} \cdot \tan(\varphi_{t}) \cdot \gamma  \cdot \Delta t \\
v_{t+1} & =  \beta \cdot v_{t} +  p_{t} \cdot \Delta t 
\end{split}
\end{equation}
where, $\beta$ and $\gamma$ are tuneable hyperparameters during optimization. The assumption of no-slip condition is valid under low or moderate vehicle velocity $v$ when making turns. As described in Subsection \ref{sec:new_model_training}, we introduce an additional cost function into the optimization to ensure that the vehicle does not overshoot its speed limit during navigation. Additionally, the steering angle is restricted to a maximum of $45^\circ$. This constrains the turn rate of the vehicle and is taken into consideration when determining the steering commands during the optimization process, also described in Subsection \ref{sec:new_model_training}. Furthermore, in Section \ref{sec:discussion}, we demonstrate that our model is robust to imprecise vehicle kinematics resulting from disturbances caused by noise in the steering command.

\subsection{Label Generation and Model Training}\label{sec:new_model_training}

An optimization-based procedure is used to determine the ground truth steering labels. The cost function to be minimized as part of the optimization comprises 4 parts: The target cost $C_{tar}$, the obstacle $C_{col\_obs}$ and vehicle $C_{col\_veh}$ collision costs, and finally the velocity cost $C_{vel}$. This cost function is designed to take the following factors into consideration: \textit{Reliability, Safety, and Speed limits.} \\

\noindent{\textbf{{Reliability:}}}
For a well-functioning model, it is imperative that all vehicles are capable of reliably reaching the desired destination state. To ensure this reliability, we introduce the target cost ($C_{tar}$), whose purpose is to derive actions such that the difference between the current and the target state for all vehicles is minimized. Mathematically:
\begin{equation}
\begin{split}
C_{tar} & =  \sum^{H}_{t=1} \sum^{N_{vehicle}}_{i=1} \| X^{(i)}_{t} -  X^{(i)}_{target} \|_{2} \cdot w_{pos} \\ 
& + \| \theta^{(i)}_{t} - \theta^{(i)}_{tar}|\|_{2}\cdot w_{orient} 
\end{split}
\end{equation}

Note that $X=[x, y]^T$ represents the position vector.\\

\noindent{\textbf{Safety:}}
It is also critical that safe driving behaviour is followed by all vehicles. This can be done by ensuring that a minimum safe distance is consistently maintained between the vehicles and between a vehicle and an obstacle. For this we introduce the vehicle collision cost, $C_{col\_veh}$ :\\
\begin{equation}
\begin{split}
C_{coll\_veh} & = \sum^{H}_{t=1} \sum^{N_{vehicle}-1}_{i=1} \sum^{N_{vehicle}}_{j=i+1} [\frac{1}{(\| X^{(i)}_{t} - X^{(j)}_{t}\|_{2}} \\
& - \frac{1}{r_{mar\_veh})}] \cdot  \Pi^{i,j}_{veh} \cdot w_{col\_veh} \\
  \Pi^{i,j}_{veh} & =
    \begin{cases}
      1 & (\| X^{(i)}_{t} - X^{(j)}_{t}\|_{2} - r_{mar\_veh}) < 0 \\
      0 & \text{otherwise}
    \end{cases} 
\end{split}
\end{equation}

and the obstacle collision cost,  $C_{col\_obs}$ :

\begin{equation}
\begin{split}
C_{coll\_obs} & = \sum^{H}_{t=1} \sum^{N_{vehicle}}_{i=1} \sum^{N_{obstacle}}_{j=1} [\frac{1}{\| X^{(i)}_{t} - X^{(j)} \|_{2} 
- r^{(j)}} \\ & - \frac{1}{r_{mar\_obs}} ]\cdot \Pi^{i,j}_{obs} \cdot w_{col\_obs} \\
  \Pi^{i,j}_{obs} & =
    \begin{cases}
      1 & (\| X^{(i)}_{t} - X^{(j)} \|_{2} - r^{(j)} - r_{mar\_obs}) < 0)\\
      0 & \text{otherwise}
    \end{cases}  
\end{split}
\end{equation}
 
Both these collision costs penalize the situation when a vehicle is too close to either another vehicle or an obstacle within a threshold safety margin ($r_{mar\_obs}$ for an obstacle and $r_{mar\_veh}$ for a vehicle). Therefore, a penalty is incurred not only when a vehicle $i$ collides with vehicle or obstacle $j$ but also when it is in its vicinity with a distance margin less than $r_{mar\_veh}$ or $r_{mar\_obs}$. This means that once a vehicle enters the safety margin, the cost increases in inverse proportion to the distance between the vehicles/obstacle.\\ Note that the obstacle can be of arbitrary shape. To further ensure safety and conservative behavior when passing by obstacles, the optimization treats each obstacle as the smallest circle of radius $r$, which entirely encapsulates the polygonal obstacle of arbitrary shape. An added advantage is that the circular representation is computationally expedient for optimization.\\

\noindent{\textbf{{Speed Limits:}}}
It is also crucial that the velocity constraints of the kinematic model and environment are respected. The velocity cost ($C_{vel}$) penalizes when vehicles tend to drive too fast \& overshoot the desired speed limit. For a vehicle with velocity $v$, a penalty is introduced if the velocity is out of a margin of $v_{mar}$. 
\begin{equation}
\begin{split}
C_{vel} & = \sum^{H}_{t=1} \sum^{N_{vehicle}}_{i=1} max(|v_{i}| - v_{mar}, 0) \cdot w_{vel} \\
\end{split}
\end{equation}

Note that $w_{pos}$, $w_{orient}$, $w_{col\_obs}$, $w_{col\_veh}$ and $w_{vel}$ are the tunable weights. A weighted sum of these cost functions is minimized to determine the ground truth control commands of each vehicle for $H$ timesteps ahead. A sequential least square programming approach is used for cost minimization \cite{kraft1988software}. \\

Note that when optimizing for the ground truth labels of hard samples, we already have the predictions from the pre-trained GNN model. Although not perfect, these predictions should be closer to the ground truth labels. They are, therefore, used to initialize the optimization. The experimental results show that this process  is approximately up to 1.8 times faster than the method that initializes the optimization from the previous timestep \cite{multiagent2023} instead of using prediction of the pre-trained model

Both the pre-trained and our improved GNN models are trained by minimizing the MSE Loss between the predicted and ground truth labels. The interested reader may refer to the code on the project page for the parameter values selected for optimization, the architecture of the GNN, and the complete implementation of the label generation and training process.\footnote{https://yininghase.github.io/multiagent-collision-mining/}.

\section{EXPERIMENTS} \label{sec:experiments}

\begin{table*}[h]
\centering
\resizebox{\textwidth}{!}{
 \begin{tabular}{||c c c c c c c c c c c||} 
 \hline
 \hline
  Number of & Number of & \multicolumn{1}{|c|}{Baseline} &\multicolumn{7}{c}{Baseline Model trained with Hard Data} & \multicolumn{1}{|c||}{Baseline Model trained} \\
  Vehicles & Obstacles & \multicolumn{1}{|c|}{Model \cite{multiagent2023}}  & 10\% & 20\% & 30\%  & 40\% & 50\%  & 60\% & 100\% (Ours) & \multicolumn{1}{|c||}{with Additional Data} \\ [0.5ex] 
 \hline
 \multicolumn{11}{||c||}{success-to-goal rate $\uparrow$ } \\ 
 \hline
8 & 0 & 0.8227 & 0.8874 & 0.8585 & 0.8988 & 0.8799 & 0.8882 & 0.8926 & 0.8959 & 0.8589 \\
\hline
8 & 1 & 0.8103 & 0.8807 & 0.8652 & 0.8937 & 0.8886 & 0.8912 & 0.8998 & 0.8857 & 0.8684 \\
\hline
10 & 0 & 0.7007 & 0.8117 & 0.7745 & 0.8399 & 0.8165 & 0.8140 & 0.8165 & 0.8342 & 0.7735 \\
\hline
10 & 1 & 0.6938 & 0.8078 & 0.7859 & 0.8350 & 0.8258 & 0.8243 & 0.8256 & 0.8126 & 0.7792 \\
\hline
12 & 0 & 0.5806 & 0.7335 & 0.7042 & 0.7632 & 0.7353 & 0.7298 & 0.7318 & 0.7581 & 0.6679 \\
\hline
12 & 1 & 0.5604 & 0.7283 & 0.7010 & 0.7702 & 0.7564 & 0.7452 & 0.7402 & 0.7379 & 0.6878 \\
\hline
15 & 0 & 0.3688 & 0.5725 & 0.5366 & 0.6455 & 0.6051 & 0.5881 & 0.5784 & 0.6234 & 0.4892 \\
\hline
20 & 0 & 0.1552 & 0.3518 & 0.3488 & 0.4165 & 0.3981 & 0.3228 & 0.3367 & 0.3782 & 0.2124 \\
\hline
\multicolumn{11}{||c||}{collision rate $\downarrow$ } \\ 
\hline
8 & 0 & 2.2904E-03 & 1.4147E-03 & 1.8451E-03 & 1.3009E-03 & 1.5388E-03 & 1.4259E-03 & 1.4547E-03 & 1.3816E-03 & 1.9673E-03 \\
\hline
8 & 1 & 2.5381E-03 & 1.5088E-03 & 1.7866E-03 & 1.3582E-03 & 1.4155E-03 & 1.3747E-03 & 1.3095E-03 & 1.5867E-03 & 1.8183E-03 \\
\hline
10 & 0 & 3.8767E-03 & 2.3360E-03 & 2.8822E-03 & 2.0089E-03 & 2.2848E-03 & 2.3737E-03 & 2.4625E-03 & 2.1422E-03 & 3.1224E-03 \\
\hline
10 & 1 & 3.9457E-03 & 2.3969E-03 & 2.7547E-03 & 2.0673E-03 & 2.1366E-03 & 2.2461E-03 & 2.2986E-03 & 2.5223E-03 & 3.0377E-03 \\
\hline
12 & 0 & 5.4523E-03 & 3.3151E-03 & 3.7751E-03 & 2.9753E-03 & 3.2651E-03 & 3.4060E-03 & 3.6316E-03 & 3.0812E-03 & 4.6445E-03 \\
\hline
12 & 1 & 5.8102E-03 & 3.4331E-03 & 3.8187E-03 & 2.8510E-03 & 2.9622E-03 & 3.2155E-03 & 3.4992E-03 & 3.5310E-03 & 4.3918E-03 \\
\hline
15 & 0 & 8.4092E-03 & 5.3944E-03 & 5.9064E-03 & 4.3712E-03 & 4.7450E-03 & 5.1751E-03 & 5.8731E-03 & 4.6981E-03 & 7.4700E-03 \\
\hline
20 & 0 & 1.3180E-02 & 8.9702E-03 & 8.6857E-03 & 7.5399E-03 & 7.7287E-03 & 9.9601E-03 & 1.0682E-02 & 8.3417E-03 & 1.4258E-02 \\
\hline
\end{tabular}
}
\caption{Shows the performance of the Baseline model (first column), Our model (second last column) and the Baseline model trained with additional random data (last column). The remaining columns in the table show the result of our model trained with progressively increasing the amount of training data. The performance is measured using the \textit{success-to-goal}  and \textit{collision rate} metrics. As more data is added to the training set, the performance of the model shows an increasing trend for the success-to-goal and a decreasing trend for the collision rate. The trend is more conspicuous in the early stages when the most hardest samples are added to the training set. The performance then tends to saturate as the less harder examples are added. It is interesting to note that our method already outperforms the other two methods with just 10\% of data. }
\label{table:statistic evaluation of 7 models}
\vspace*{-3mm}
\end{table*}

In this section we give an overview of the models, evaluation metrics, quantitative results, while also providing a discussion on the interpretation of the results. 
\subsection{Models}\label{subsec:models}
For the purpose of comparison, we additionally evaluate the following models:\\

\noindent{\textbf{Baseline \cite{multiagent2023}}}
This is the model from \cite{multiagent2023}. It did not mine for hard samples when determining the labels. The optimization initializes the control values from the previous timestep rather than the model itself. It was also trained only with up to 3 vehicles and 0 obstacles and was therefore used as the pre-trained model. \\

\noindent{\textbf{Baseline \cite{multiagent2023} with additional (random) data:}} This model is similar to the previous baseline model except that it was additionally trained on data with up to 4 vehicles and 1 obstacle. As is the case with the baseline, this model was also trained with samples selected randomly, without consideration for how easy or difficult they are for the pre-trained model. Another difference from the baseline is that the optimization of the new data was initialized with predictions from the pre-trained model. \\

In both our method and the baseline method with additional data, we further include 50 trajectories for each of the following scenarios: 3 vehicles 0 obstacle, 3 vehicle 1 obstacle and 3 vehicle 2 obstacles. Meanwhile, 40 trajectories are added for 4 vehicles 0 obstacles and 4 vehicle 1 obstacle. Each trajectory contains 120 data points.

\begin{figure*}[ht]
\centering
    \subfigure[\centering
    {Our Model}]{{\includegraphics[width=0.30\textwidth]{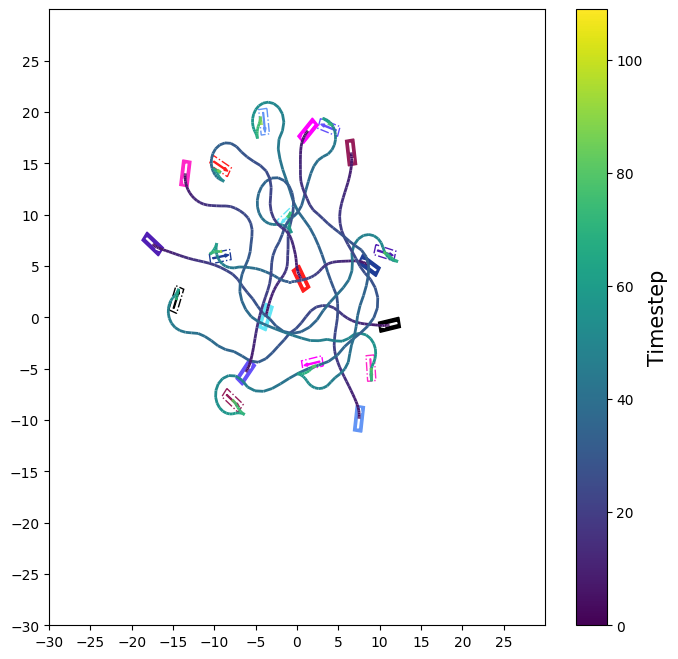}}}
    \subfigure[\centering
    {Baseline \cite{multiagent2023} with additional data}]{{\includegraphics[width=0.30\textwidth]{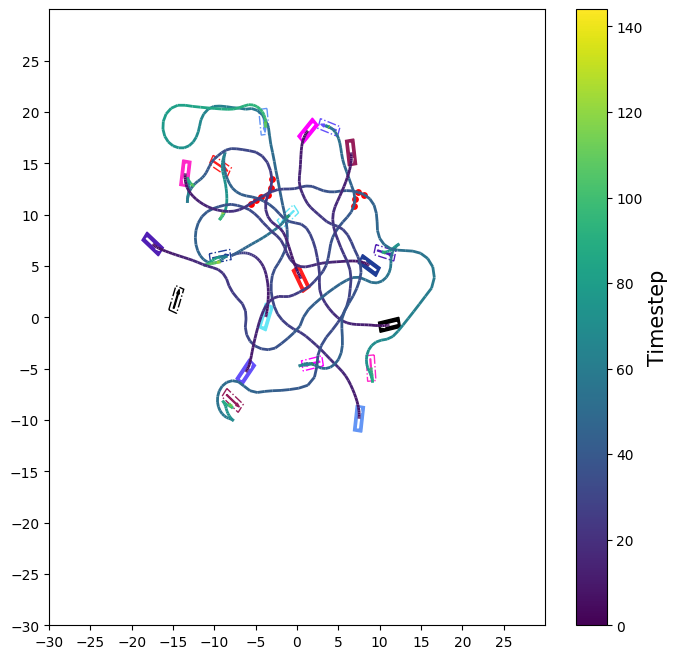}}}
    \subfigure[\centering
    {Baseline \cite{multiagent2023} }]{{\includegraphics[width=0.30\textwidth]{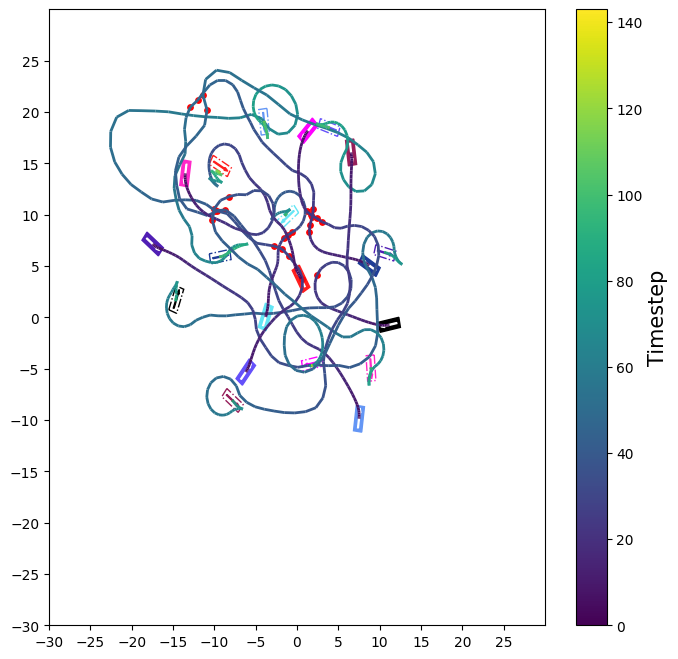}}}
    \quad

    \caption{Shows a qualitative visualization of the trajectory traversed by the 3 models for a sample scenario. The red dots on the trajectories indicate collision between the vehicles. Only our model does not have collisions, while the other two have plenty. Corresponding videos can be visualized on the project page: \url{https://yininghase.github.io/multiagent-collision-mining/\#Comparison-with-Other-Models}}
    \label{fig:compare_of_3_models}
\end{figure*}

\subsection{Testing Environment:}
Testing the models for vehicle navigation involves the interaction of the individual agents with the environment. This necessitates closed-loop online evaluation. Although there are plenty of driving datasets such as \cite{Geiger2012CVPR,nuscenes2019}, they only cater to single vehicles. Moreover, even if these datasets could be extended to multiple vehicles, they are limited to the evaluation of perception-related tasks. Therefore, in order to assess the online performance of vehicle control tasks, simulation engines have proven to be expedient. They have accelerated research in this direction by allowing different algorithms proposed in literatures to be benchmarked against a common framework \cite{carla-leaderboard}. These simulation environments have successfully been used for the evaluation of both single and multi-agent control algorithms \cite{codevilla2018end,ZHU2024127220}. For our task, we adapt the environment from \cite{multiagent2023}. 

\subsection{Evaluation Metrics:}

We use the \textit{success-to-goal} and \textit{collision rate} metrics for evaluation. The \textit{success-to-goal rate} is the ratio of vehicles that are capable of successfully reaching their target destination within a certain threshold without colliding with obstacles or any other vehicles in the scene. The threshold was set as 0.2 radians for the orientation and 1.25m for the difference between the location of the vehicle and its destination state. Meanwhile, the \textit{collision rate} is the ratio of the total number of collisions to the total distance traveled by all vehicles in a scenario. A higher value for the success-to-goal and a lower value for collision rate is better.

We additionally report the \textit{runtime} metric. It is the amount of time (in seconds) it takes to execute one step of the optimization process. 

\subsection{Quantitative Results:}
Performance of the models described in Subsection \ref{subsec:models} for scenarios comprising of different numbers of vehicles/obstacles in the scene is enumerated in Table \ref{table:statistic evaluation of 7 models}. To enhance the challenge, we emphasize increasing the number of vehicles rather than obstacles in the scene. This is because each additional vehicle can be considered a dynamic obstacle for the other vehicles already on the scene. To thoroughly evaluate the models, we test on a large number of diverse trajectories. Each row/scenario in the table comprises 4000 test samples. Since there are 8 different scenarios in the table, this amounts to 32,000 test trajectories in total. It is pertinent to note that none of the vehicle/obstacle combinations were present in the training set. The training data had a maximum of 4 vehicles, while the test set comprised at least 8 vehicles. As can be seen, our method outperforms both the baseline models for all the scenarios on both the success-to-goal and collision rate metrics.

Meanwhile, Table \ref{table:statistic evaluation of 7 models} also shows how the performance of our model evolves as hard samples based on the collision rate metrics are progressively added to the training. It can be observed that our model achieves better performance than both baseline models with only a fraction of the samples. 

\begin{figure*}[!ht]
\centering
    \subfigure[\centering
    {8 vehicles, 0 obstacle}]{{\includegraphics[width=0.24\textwidth]{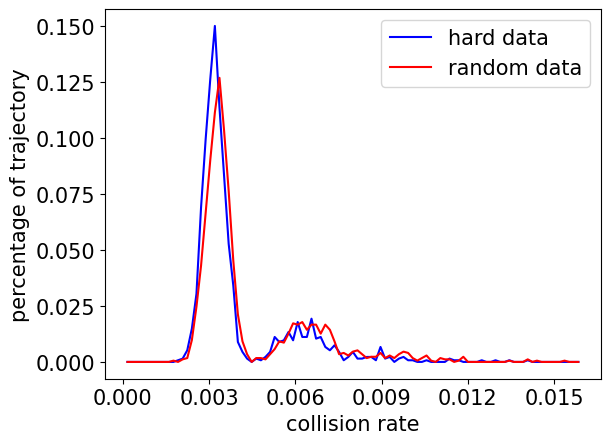}}}
    \subfigure[\centering
    {12 vehicles, 0 obstacle}]{{\includegraphics[width=0.24\textwidth]{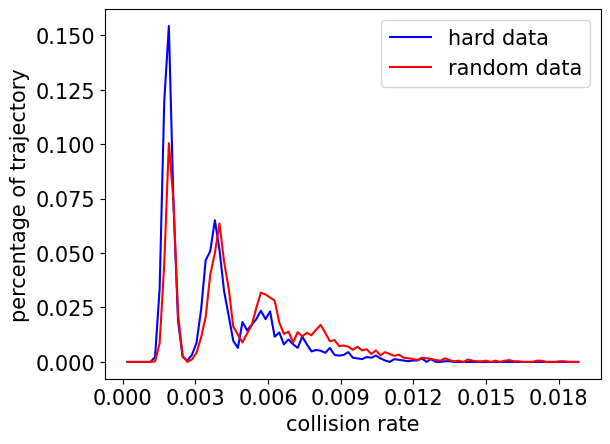}}}
    \subfigure[\centering
    {15 vehicles, 0 obstacle}]{{\includegraphics[width=0.24\textwidth]{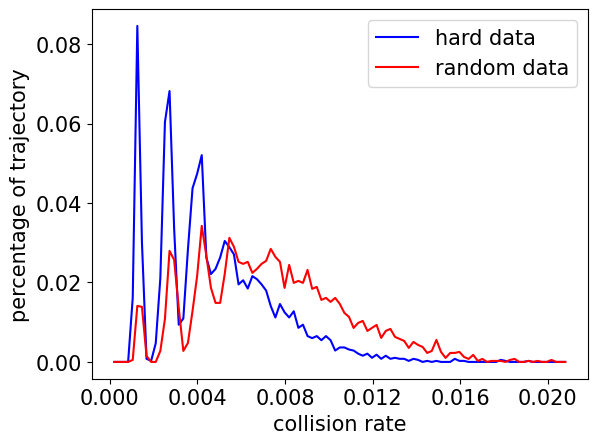}}}
    \subfigure[\centering
    {20 vehicles, 0 obstacle}]{{\includegraphics[width=0.24\textwidth]{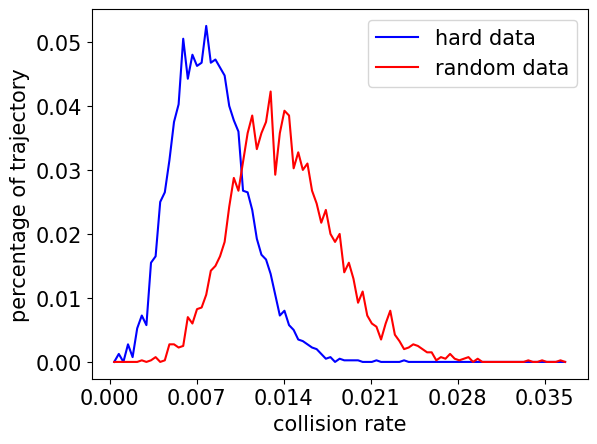}}}
    \quad
    \caption{Shows the normalized probability density function of trajectories at different collision rates for 4 vehicle/obstacle configurations.  It can be observed that the distribution of the baseline method trained with additional random data (red curve) is always to the right of the distribution of our method (blue curve). This is more pronounced as the number of vehicles  in the scene are increased. This shows that the method trained with random samples has more occurrences with higher collision rates.  }
    \label{fig:collision_rate_histogram_of_models}
    \vspace*{-4mm}
\end{figure*}

Figure \ref{fig:collision_rate_histogram_of_models} shows the normalized probability density function of trajectories for different collision rates for both the baseline model with additional data and our method. Note that only trajectories with collisions are kept in the visualization. One can see that the distribution for the random sampling method is shifted to the right of the distribution produced by our method. This shows that for the baseline method with additional data, there tend to be more samples with a higher collision rate metric.

Also, a qualitative visualization of the trajectories traversed by the 3 models for a sample scenario is provided in Figure \ref{fig:compare_of_3_models}. The optimization runtimes for the different configurations are given in Table \ref{table:statistic of optimization run time}. Lastly, the performance of the baseline and our model in response to disturbances induced in the steering angle resulting from imprecise vehicle kinematics is given in Table \ref{table: statistic of absolute decreased performance on steering angle noise}. 

\subsection{Discussion}\label{sec:discussion}

\noindent{\textbf{Comparison:}} Table \ref{table:statistic evaluation of 7 models} shows the results of all the models described in Subsection \ref{subsec:models}. As can be observed, our method outperforms both the other models for all the scenarios. Outperforming the baseline model is not surprising since it was only trained with up to 3 vehicles and 0 obstacles, whereas our model was trained with up to 4 vehicles and 1 obstacle. However, what is intriguing is that our model also has superior performance to the model trained with random sampling across all the scenarios. This is despite the fact that the baseline model with additional random samples was also trained with up to 4 vehicles and 1 obstacle and on the same amount of data. The only difference is that we mined for samples which had the highest collision rate metric and were therefore difficult for the pre-trained baseline model to make the correct prediction. Therefore, the performance gain for our model can be attributed to this mining for hard samples. This leads to a further line of investigation: Can we use a lesser number of "hard" samples and achieve at-par performance? We address this in the next point.\\

\noindent{\textbf{Progressive Hard Sample Mining: }} To better investigate how mining for hard samples influences the performance of the model, we first rank the samples in order of difficulty. The difficulty is based on the collision rate metric, with higher values of the metric alluding to greater difficulty. Based on this ranking, we successively keep adding 10\% of the most difficult samples for training until we reach 60\%, where the performance has saturated closer to the final one with 100\% of the samples. The results show a precipitous enhancement in performance when including the initial samples, so much so that the first 10\%  already outperforms the model trained with the random sampling method for all the scenarios. Successive inclusion of the samples progressively enhances performance but at a slower rate until saturation.
 
This shows that the tedious time-consuming label generation process can be reduced 10-fold to achieve at-par performance by resourcefully selecting only those samples for training that are potentially difficult for the model. \\
 
 \noindent{\textbf{Error Distribution: }} The numbers in Table \ref{table:statistic evaluation of 7 models} only report the mean statistic of the collision rates of the models across the different scenarios. A better metric would be to report the distribution of the collision rates to better assess how our method compares with random sampling. Figure \ref{fig:collision_rate_histogram_of_models} shows a probability density function of the trajectories for different collision rates. It can be observed that the distribution for the random sampling method tends to usually be towards the right of our method. This demonstrates that the random sampling method has more trajectories with higher collision rates than our method. It can also be observed that for scenarios with fewer vehicles, the difference between the distributions is minimal, which progressively tends to become more noticeable with a higher number of vehicles. This demonstrates that our approach of mining hard samples becomes more favorable for scenarios having a larger number of vehicles at inference time. \\
    
\noindent{\textbf{Runtime Optimization:}} Table \ref{table:statistic of optimization run time} shows the time taken by the optimization to converge to a solution. Comparison is made between our method ,which uses model predictions to initialize the control values to be optimized, and the approach from the baseline method, which utilizes the predictions from the previous timestep. It can be seen from the table that our method is up to 1.8 times faster. One plausible explanation is that the model predictions are already close to the optimal solutions and hence require less iterations to converge.

Note that the runtime performance for all cases in Table \ref{table:statistic of optimization run time} were evaluated on a machine with an Intel Core i7-10750H CPU and GeForce RTX 2070 GPU.\\

\begin{table}
\centering
\resizebox{\linewidth}{!}{
 \begin{tabular}{||c c c c c||} 
 \hline
 Number of & Number of & Optimization without & Optimization with & ratio without/with\\ 
 Vehicle & Obstacle & GNN initialization \cite{multiagent2023} & GNN initialization (Ours) & GNN initialization\\ 
 [0.5ex] 
 \hline\hline
1 & 0 & 0.63230 & 0.39116 & 1.616 \\
\hline
1 & 1 & 0.73012	& 0.46215 & 1.580 \\
\hline
1 & 2 & 0.74968 & 0.46262 & 1.621 \\
\hline
1 & 3 & 0.75063 & 0.50808 & 1.477 \\
\hline
1 & 4 & 0.79725 & 0.49777 & 1.602 \\
\hline
2 & 0 & 2.74806 & 1.58956 & 1.729 \\
\hline
2 & 1 & 3.10209 & 1.85043 & 1.676 \\
\hline
2 & 2 & 3.06539 & 1.77575 & 1.726 \\
\hline
2 & 3 & 2.96316 & 1.72915 & 1.714 \\
\hline
2 & 4 & 2.96338 & 1.66151 & 1.784 \\
\hline
3 & 0 & 5.48114 & 3.01684 & 1.817 \\
\hline
3* & 1* & 5.75448 & 3.77693 & 1.524 \\
\hline
3* & 2* & - & 3.70428 & - \\
\hline
3* & 3* & - & 3.91798 & - \\
\hline
3* & 4* & - & 3.71874 & - \\
\hline
4* & 0* & - & 5.15936 & - \\
\hline
4* & 1* & - & 5.86252 & - \\
\hline
\end{tabular}
}
\caption{Shows the average run time (in secs) per optimization step with and without the pre-trained GNN model prediction as initialization. It can be observed that our approach, which uses initialization of the GNN model for optimization, is up to 1.8 times faster than the method, which uses initialization values from the previous timestep. The rows with the asterisk are those that were additionally included in our training samples and only used the faster method for optimization. Therefore, the runtimes for the slower optimization method are not reported except for the scenario with 3 vehicles and 1 obstacles, which is around 1.5 times slower than our method.}
\label{table:statistic of optimization run time}
\vspace*{-4mm}
\end{table}

\noindent{\textbf{Robustness to Noise:}} Note that, we use the vehicle kinematic model to determine the hard negative samples without actually having to deploy the model in real-time. However, one may argue that this modeling may not perfectly reflect the real world. Furthermore, imperfections in the kinematic modeling have the potential to introduce disturbances in the navigation of vehicles at inference time. Hence, the steering command predicted by the GNN model may not lead to the desired action being executed.  Ideally, a robust model should be less sensitive to these disturbances and successfully drive all the agents to the target without many collisions. 

Therefore, to assess the robustness of a GNN model in face of such disturbances, we inject noise into the steering angle predicted by the model at progressively increasing levels of intensity. The influence of this noise on the performance of the model is then measured.

\begin{figure}[ht]
  \centering
  \includegraphics[width=\linewidth]{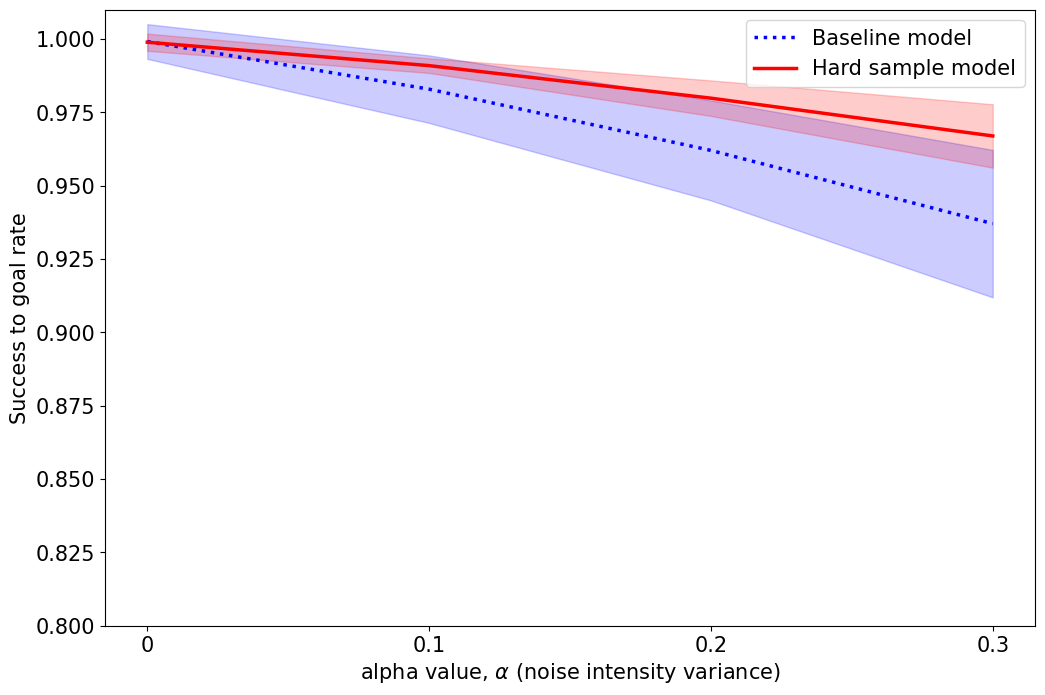} 
  \caption{Shows the relative change in \textit{success-to-goal} ratio when the variance of noise intensity controlled by $\alpha$ is increased from 0 to 0.3. All values are normalized in reference to the results for  $\alpha$ = 0.  The curves show the mean performance across all the vehicle/obstacle configurations reported in Table \ref{table: statistic of absolute decreased performance on steering angle noise} for the baseline and our model trained on hard sampled data. The figure also depicts the standard deviation for each model, as shown by the shaded regions. The results in this figure show that our model is not only relatively more robust to noise in the steering angle as depicted by a slower drop in \textit{success-to-goal} ratio but also is more stable in its prediction across the different configurations as demonstrated by a lower standard deviation.  }
  \label{fig: statistic of relative decreased performance on steering angle noise}
  \vspace*{-6mm}
\end{figure}

The noise $\Delta \varphi$  added to the steering angle is sampled from a zero mean Gaussian with variance proportional to the absolute value of the steering angle (scaled by $\alpha$)  plus a bias term $\beta$. Mathematically: $\Delta \varphi \sim{N(0, \alpha * |\varphi| + \beta)}$. In our setting, $\beta$ is fixed to be $2^\circ$.  Table \ref{table: statistic of absolute decreased performance on steering angle noise} shows the consequence of increasing the  $\alpha$ value from 0 to 0.3 in steps of 0.1 on the model performance. The results show that our model trained with hard samples is less sensitive to noise when compared with the baseline model, whose success-to-goal rate metric decline is relatively steeper. This is more apparent in Figure \ref{fig: statistic of relative decreased performance on steering angle noise}, which shows the relative change in performance from the no noise level.

For those interested, the project page also contains experiments due to noise induced in the vehicle state from imprecise sensor measurements, runtimes for GNN inference along with figures for the error distribution for all configurations.
\begin{table*}[!h]
\centering
\resizebox{\textwidth}{!}{
 \begin{tabular}{||c c c c c c c c c c c c||} 
 \hline
 \hline
  Number of & Number of & \multicolumn{5}{|c|}{Baseline Model \cite{multiagent2023}} &\multicolumn{5}{c||}{Baseline Model trained with Hard Samples} \\
  Vehicles & \multicolumn{1}{c|}{Obstacles} & no noise & 0 & 0.1 & 0.2 & \multicolumn{1}{c|}{0.3}  & no noise & 0 & 0.1 & 0.2 & 0.3 \\ [0.5ex]
  \hline
  8 & 0 & 0.8227 & 0.8284 & 0.8213 & 0.8136 & 0.8022 & 0.8959 & 0.8951 & 0.8884 & 0.8801 & 0.8794 \\
  8 & 1 & 0.8103 & 0.8121 & 0.8063 & 0.7949 & 0.7819 & 0.8857 & 0.8821 & 0.8798 & 0.8711 & 0.8633 \\
  10 & 0 & 0.7007 & 0.7035 & 0.6944 & 0.6793 & 0.6661 & 0.8342 & 0.8363 & 0.8264 & 0.8231 & 0.8067 \\
  10 & 1 & 0.6938 & 0.6906 & 0.6804 & 0.6636 & 0.6468 & 0.8126 & 0.8116 & 0.8059 & 0.8024 & 0.7935 \\
  12 & 0 & 0.5806 & 0.5743 & 0.5652 & 0.5561 & 0.5331 & 0.7581 & 0.7556 & 0.7511 & 0.7376 & 0.7321 \\
  12 & 1 & 0.5604 & 0.5621 & 0.5518 & 0.5353 & 0.5203 & 0.7379 & 0.7341 & 0.7284 & 0.7208 & 0.7123 \\
  15 & 0 & 0.3688 & 0.3681 & 0.3586 & 0.3451 & 0.3351 & 0.6234 & 0.6241 & 0.6158 & 0.6056 & 0.5961 \\
  20 & 0 & 0.1552 & 0.1544 & 0.1501 & 0.1477 & 0.1417 & 0.3782 & 0.3791 & 0.3761 & 0.3697 & 0.3588  \\

  \hline
  \hline
  \end{tabular}
}
\caption{Comparison of the \textit{success-to-goal} rate  performance of the baseline mode with our model trained on hard samples. The performance is reported for increasing values of noise variance intensity  $\alpha$ increasing from 0 to 0.3 in steps of 0.1. It can be seen that our model trained with hard samples is less sensitive to noise intensity than the baseline model. }
\label{table: statistic of absolute decreased performance on steering angle noise}
\vspace*{-5mm}
\end{table*}

\section{Conclusion}\label{sec:conclusion}
In this paper, we demonstrated how a model can be trained to control multiple vehicles in an unstructured environment without any traffic management system, such as traffic lights, road signs, etc. This is akin to driving in places where traffic management systems are non-existent, and road infrastructure is not available. We further demonstrated that our approach of mining for negative samples significantly reduces the number of samples to be labeled by a factor of 10. This can be useful in expediting the deployment of autonomous driving algorithms in such non-standardized environments.

\bibliographystyle{IEEEtran}
\bibliography{root}

\end{document}